\documentclass[11pt,showpacs,showkeys,preprintnumbers,amsmath,amssymb]{revtex4}

\begin{document}
\title{Established pseudo solution of second-order Dirac-Coulomb equation
with position-dependent mass}
\author{Ruida Chen³Âï}

\affiliation{Shenzhen Institute of Mathematics and Physics,
Shenzhen, 518028, China}


\begin{abstract}
We show that one of exact solutions of the second-order
Dirac-Coulomb equation were pseudo. In the corresponding original
literature, it was considered that the mass of the electron with a
Coulomb potential was position-dependent, but the obtained
eigenvalues set was not the inevitable mathematical deduction of the
given second-order Dirac equation, and the second-order Dirac
equations were not the inevitable mathematical deduction of the
given couplet first-order Dirac equation with the position-dependent
mass of the electron. In the present paper, we obtain the correct
solution of the introduced first-order differential equations. This
new solution would be tenable only when the wave equation is
correct, but there is not any experiment date to validate the
so-called position-dependent of the electron in the Coulomb field.
\end{abstract}



\maketitle

\section{Introduction}

An exact solution of the Dirac equation with the so-called
position-dependent mass of the electron in the Coulomb field was
introduced\cite{Alhaidari:2004}. According to the article, in atomic units
($\hbar = m_0 = 1)$, the spherically symmetric singular mass distribution
was taken as follows
\begin{equation}
\label{eq1}
m\left( r \right) = 1 + {\mu \lambda ^2} \mathord{\left/ {\vphantom {{\mu
\lambda ^2} r}} \right. \kern-\nulldelimiterspace} r
\end{equation}

\noindent
where $\lambda $ is the Compton wavelength $\hbar \mathord{\left/ {\vphantom
{\hbar {m_0 c}}} \right. \kern-\nulldelimiterspace} {m_0 c} = c^{ - 1}$, and
$\mu $ is a real scale parameter with inverse length dimension. The author
presented some reasons why he introduced the position-dependent mass of the
electron. We find the reasons were independent of any physical logic. For
example, it was of that the rest mass of the particle $\left( {m_0 = 1}
\right)$ was obtained either as the asymptotic limit ($r \to \infty )$, or
the nonrelativistic limit ($\lambda \to 0)$ of $m\left( r \right)$,
consequently, a possible interpretation for this singular mass term might be
found in relativistic quantum field theory. It was even told of that it
should also be noted that this position-dependent mass term has a
relativistic origin as well since it was proportional to the Compton
wavelength which vanishes as $c \to \infty $(equivalently, $\lambda \to 0)$.

Whereas we firstly query a question here, did any experiment ever show that
the mass of electron in the Coulomb field is relative to the position? One
can also find that many such suppositions in the corresponding published
paper are very vexed. We are clear that there is not any consequence for the
$m_0 $ and relativistic result. Of course, at present, we have to
transitorily avoid such questions and only check the corresponding
mathematic deduction procedure. We show that the original solution of the
second-order Dirac equation with the so-called position-dependent mass was
incorrect, and the given second-order Dirac equation is not the necessary
mathematical deduction of the given first-order Dirac equation with
position-dependent mass of the electron in the Coulomb field. We introduce
the correct exact solution of the original first-order differential equation
only for further showing that the original solution includes many
mathematical mistakes, and don't think the supposition of the
position-dependent mass of the electron in the Coulomb field is correct.

\section{Established solution of Dirac equation with position-dependent mass}

In order to solve the Dirac equation with the so-called position-dependent
mass term in the Coulomb field, the spinor wavefunction was written as
follows
\begin{equation}
\label{eq2}
\psi = \left( {{\begin{array}{*{20}c}
 {i\left[ {{g\left( r \right)} \mathord{\left/ {\vphantom {{g\left( r
\right)} r}} \right. \kern-\nulldelimiterspace} r} \right]\chi _{\iota m}^j
} \hfill \\
 {\left[ {{f\left( r \right)} \mathord{\left/ {\vphantom {{f\left( r
\right)} r}} \right. \kern-\nulldelimiterspace} r} \right]\vec {\sigma }
\cdot \chi _{\iota m}^j } \hfill \\
\end{array} }} \right)
\end{equation}

\noindent
where $f$ and $g$ are real radial functions, $\hat {r}$ is the radial unit
vector, and the angular wavefunction with the spherical harmonic function
$Y_\iota ^{m - 1 \mathord{\left/ {\vphantom {1 2}} \right.
\kern-\nulldelimiterspace} 2} $ for the two-component spinor was written as
\begin{equation}
\label{eq3}
\chi _{\iota m}^j = \frac{1}{\sqrt {2\iota + 1} }\left(
{{\begin{array}{*{20}c}
 {\sqrt {\iota \pm m + 1 \mathord{\left/ {\vphantom {1 2}} \right.
\kern-\nulldelimiterspace} 2} } \hfill & {Y_\iota ^{m - 1 \mathord{\left/
{\vphantom {1 2}} \right. \kern-\nulldelimiterspace} 2} } \hfill \\
 { \mp \sqrt {\iota \mp m + 1 \mathord{\left/ {\vphantom {1 2}} \right.
\kern-\nulldelimiterspace} 2} } \hfill & {Y_\iota ^{m + 1 \mathord{\left/
{\vphantom {1 2}} \right. \kern-\nulldelimiterspace} 2} } \hfill \\
\end{array} }} \right)
\end{equation}

\noindent
then the following $2\times 2$ matrix equation for the two radial spinor
components was given
\begin{equation}
\label{eq4}
\left( {{\begin{array}{*{20}c}
 {1 + \lambda ^2\frac{Z + \mu }{r} - \varepsilon } \hfill & {\lambda \left(
{\frac{k}{r} - \frac{d}{dr}} \right)} \hfill \\
 {\lambda \left( {\frac{k}{r} + \frac{d}{dr}} \right)} \hfill & { - 1 +
\lambda ^2\frac{Z - \mu }{r} - \varepsilon } \hfill \\
\end{array} }} \right)\left( {{\begin{array}{*{20}c}
 {g\left( r \right)} \hfill \\
 {f\left( r \right)} \hfill \\
\end{array} }} \right) = 0
\end{equation}

\noindent
where $\varepsilon $ is the relativistic energy which is real. By
introducing some signs and using the global unitary transformation, it was
alleged that, the Schrodinger-like wave equation was finally obtained. The
main steps are as follows. The Schrodinger-like requirement dictates that
the parameter $\eta $ satisfies the constraint
\begin{equation}
\label{eq5}
C\mu + {S\kappa } \mathord{\left/ {\vphantom {{S\kappa } {\lambda = \pm Z}}}
\right. \kern-\nulldelimiterspace} {\lambda = \pm Z}
\end{equation}

\noindent
where $S = \sin \left( {\lambda \eta } \right)$, $C = \cos \left( {\lambda
\eta } \right)$ and $ - \pi \mathord{\left/ {\vphantom {\pi 2}} \right.
\kern-\nulldelimiterspace} 2 \le \lambda \eta \le \pi \mathord{\left/
{\vphantom {\pi 2}} \right. \kern-\nulldelimiterspace} 2$, $\kappa = \pm
1,\;\pm 2,\; \cdots $. The solution of the constraint (\ref{eq5}) gives two angles
whose cosines are
\begin{equation}
\label{eq6}
C = \left( {\mu ^2 + {\kappa ^2} \mathord{\left/ {\vphantom {{\kappa ^2}
{\lambda ^2}}} \right. \kern-\nulldelimiterspace} {\lambda ^2}} \right)^{ -
1}\left[ {\pm \mu Z + \frac{\left| \kappa \right|}{\lambda }\sqrt {\left(
{\frac{\kappa }{\lambda }} \right)^2 + \mu ^2 - Z^2} } \right] > 0
\end{equation}

\noindent
the equation (\ref{eq4}) is now transformed into the following
\begin{equation}
\label{eq7}
\left( {{\begin{array}{*{20}c}
 {C - \varepsilon + \left( {1\pm 1} \right)\lambda ^2\frac{Z}{r}} \hfill &
{\lambda \left( { - \frac{s}{\lambda } + \frac{\gamma }{r} - \frac{d}{dr}}
\right)} \hfill \\
 {\lambda \left( { - \frac{s}{\lambda } + \frac{\gamma }{r} + \frac{d}{dr}}
\right)} \hfill & { - C - \varepsilon + \left( {1 \mp 1} \right)\lambda
^2\frac{Z}{r}} \hfill \\
\end{array} }} \right)\left( {{\begin{array}{*{20}c}
 {\phi ^ + \left( r \right)} \hfill \\
 {\phi ^ - \left( r \right)} \hfill \\
\end{array} }} \right) = 0
\end{equation}

\noindent
where $\gamma = \frac{\left| \kappa \right|}{\kappa }\sqrt {\kappa ^2 +
\lambda ^2\left( {\mu ^2 - Z^2} \right)} $ and
\begin{equation}
\label{eq8}
\left( {{\begin{array}{*{20}c}
 {\phi ^ + } \hfill \\
 {\phi ^ - } \hfill \\
\end{array} }} \right) = \left( {{\begin{array}{*{20}c}
 {\cos \frac{\lambda \eta }{2}} \hfill & {\sin \frac{\lambda \eta }{2}}
\hfill \\
 { - \sin \frac{\lambda \eta }{2}} \hfill & {\cos \frac{\lambda \eta }{2}}
\hfill \\
\end{array} }} \right)\left( {{\begin{array}{*{20}c}
 g \hfill \\
 f \hfill \\
\end{array} }} \right)
\end{equation}

\noindent
equation (\ref{eq7}) gives one spinor component in terms of the other as follows
\begin{equation}
\label{eq9}
\phi ^\pm = \frac{\lambda }{C\pm \varepsilon }\left( {\pm \frac{S}{\lambda }
\mp \frac{\gamma }{r} + \frac{d}{dr}} \right)\phi ^ \mp
\end{equation}

\noindent
whereas, the resulting Schr\"{o}dinger--like wave equation becomes
\begin{equation}
\label{eq10}
\left[ { - \frac{d^2}{dr^2} + \frac{\gamma \left( {\gamma \pm 1}
\right)}{r^2} + 2\frac{Z\varepsilon + \mu }{r} - \frac{\varepsilon ^2 -
1}{\lambda ^2}} \right]\phi ^ \mp \left( r \right) = 0
\end{equation}

\noindent
the author comparing this equation with that of the well-known
nonrelativistic Coulomb problem with constant mass and alleged that he had
new discover for the relativistic spectrum
\begin{equation}
\label{eq11}
\varepsilon _n^\iota = \left[ {1 + \left( {\frac{\lambda Z}{n + \iota + 1}}
\right)^2} \right]^{ - 1}\left[ { - \frac{\lambda ^2\mu Z}{\left( {n + \iota
+ 1} \right)^2}\pm \sqrt {1 + \lambda ^2\frac{Z^2 - \mu ^2}{\left( {n +
\iota + 1} \right)^2}} } \right]
\end{equation}

\noindent
namely
\begin{equation}
\label{eq12}
\varepsilon _n^\iota = \frac{ - \lambda ^2\mu Z\pm \sqrt {\left( {n + \iota
+ 1} \right)^4 + \lambda ^2\left( {Z^2 - \mu ^2} \right)\left( {n + \iota +
1} \right)^2} }{\left( {n + \iota + 1} \right)^2 + \left( {\lambda Z}
\right)^2}
\end{equation}

\noindent
where $\iota $ stands for either one of the four possible alternative values
in the following expression associated independently with $\phi ^\pm $ and
$\pm \kappa \ge 1$. For $\phi ^ + $
\begin{equation}
\label{eq13}
\iota \to \gamma \mbox{ or }\iota \to - \gamma - 1,\mbox{ Z} \to
\mbox{Z}\varepsilon \mbox{ + }\mu \mbox{, }E \to {\left( {\varepsilon ^2 -
1} \right)} \mathord{\left/ {\vphantom {{\left( {\varepsilon ^2 - 1}
\right)} {2\lambda ^2}}} \right. \kern-\nulldelimiterspace} {2\lambda ^2}
\end{equation}

\noindent
and for $\phi ^ - $
\begin{equation}
\label{eq14}
\iota \to \gamma - 1\mbox{ or }\iota \to - \gamma ,\mbox{ Z} \to
\mbox{Z}\varepsilon \mbox{ + }\mu \mbox{, }E \to {\left( {\varepsilon ^2 -
1} \right)} \mathord{\left/ {\vphantom {{\left( {\varepsilon ^2 - 1}
\right)} {2\lambda ^2}}} \right. \kern-\nulldelimiterspace} {2\lambda ^2}
\end{equation}

It is similar to the above procedure, writing the paper about the
relativistic quantum mechanics, many authors did not introduce their detail
operation steps on how to obtain those necessary transition equation and
their new mathematical result. They only alleged that those formulas in
their paper are necessary deduction. We don't understand what the expression
(\ref{eq9}) means?

\section{Eingevalues-set (\ref{eq12}) disobey uniqueness of solution of wave equation}

For the same quantum system, it should have only one of eigenvalues set for
any theory. The formula (\ref{eq12}) of the energy levels includes two eigenvalues
set corresponding to different definition$\iota $. It is one of the
mathematical contradictions of the articles. Consequently we cannot believe
the formula (\ref{eq12}) is the real energy eigenvalues set in the Coulomb field. It
seems that the author is not up on the method of finding the eigensolutions
of the second-order differential equations with variable coefficients. Why
don't we directly solve the second-order differential equation (\ref{eq10}) now?
Now, one can write the equation (\ref{eq10}) in the separate form
\begin{equation}
\label{eq15}
\begin{array}{l}
 \left[ { - \frac{d^2}{dr^2} + \frac{\gamma \left( {\gamma + 1}
\right)}{r^2} + 2\frac{Z\varepsilon + \mu }{r} - \frac{\varepsilon ^2 -
1}{\lambda ^2}} \right]\phi \left( r \right) = 0 \\
 \left[ { - \frac{d^2}{dr^2} + \frac{\gamma \left( {\gamma - 1}
\right)}{r^2} + 2\frac{Z\varepsilon + \mu }{r} - \frac{\varepsilon ^2 -
1}{\lambda ^2}} \right]\psi \left( r \right) = 0 \\
 \end{array}
\end{equation}

\noindent
using the asymptotic solution $\phi \left( r \right) \sim \exp \left( {{ -
r\sqrt {1 - \varepsilon ^2} } \mathord{\left/ {\vphantom {{ - r\sqrt {1 -
\varepsilon ^2} } \lambda }} \right. \kern-\nulldelimiterspace} \lambda }
\right)$, $\psi \left( r \right) \sim \exp \left( {{ - r\sqrt {1 -
\varepsilon ^2} } \mathord{\left/ {\vphantom {{ - r\sqrt {1 - \varepsilon
^2} } \lambda }} \right. \kern-\nulldelimiterspace} \lambda } \right)$
satisfying the boundary condition at $r \to 0$, we seek the formal solution
\begin{equation}
\label{eq16}
\phi \left( r \right) = e^{ - \frac{\sqrt {1 - \varepsilon _u^2 } }{\lambda
}r}u,\quad \psi \left( r \right) = e^{ - \frac{\sqrt {1 - \varepsilon _v^2 }
}{\lambda }r}v
\end{equation}

\noindent
it easily obtained that
\begin{equation}
\label{eq17}
\begin{array}{l}
 \frac{d\phi \left( r \right)}{dr} = e^{ - \frac{\sqrt {1 - \varepsilon _u^2
} }{\lambda }r}\left( {\frac{du}{dr} - \frac{\sqrt {1 - \varepsilon _u^2 }
}{\lambda }u} \right) \\
 \frac{d^2\phi \left( r \right)}{dr^2} = e^{ - \frac{\sqrt {1 - \varepsilon
_u^2 } }{\lambda }r}\left( {\frac{d^2u}{dr^2} - \frac{2\sqrt {1 -
\varepsilon _u^2 } }{\lambda }\frac{du}{dr} + \frac{1 - \varepsilon _u^2
}{\lambda ^2}u} \right) \\
 \frac{d\psi \left( r \right)}{dr} = e^{ - \frac{\sqrt {1 - \varepsilon _v^2
} }{\lambda }r}\left( {\frac{dv}{dr} - \frac{\sqrt {1 - \varepsilon _v^2 }
}{\lambda }v} \right) \\
 \frac{d^2\psi \left( r \right)}{dr^2} = e^{ - \frac{\sqrt {1 - \varepsilon
_v^2 } }{\lambda }r}\left( {\frac{d^2v}{dr^2} - \frac{2\sqrt {1 -
\varepsilon _v^2 } }{\lambda }\frac{dv}{dr} + \frac{1 - \varepsilon _v^2
}{\lambda ^2}v} \right) \\
 \end{array}
\end{equation}

\noindent
substituting (\ref{eq16}) and (\ref{eq17}) into (\ref{eq15}), it educes that
\begin{equation}
\label{eq18}
\begin{array}{l}
 \frac{d^2u}{dr^2} - \frac{2\sqrt {1 - \varepsilon _u^2 } }{\lambda
}\frac{du}{dr} - \frac{\gamma \left( {\gamma + 1} \right)}{r^2}u -
2\frac{Z\varepsilon _u + \mu }{r}u = 0 \\
 \frac{d^2v}{dr^2} - \frac{2\sqrt {1 - \varepsilon _v^2 } }{\lambda
}\frac{dv}{dr} - \frac{\gamma \left( {\gamma - 1} \right)}{r^2}v -
2\frac{Z\varepsilon _v + \mu }{r}v = 0 \\
 \end{array}
\end{equation}

\noindent
finding the power series solution of the above equations, it assumed that
\begin{equation}
\label{eq19}
u = \sum\limits_{n = 0}^\infty {b_n r^{s_u + n}} ,\quad v = \sum\limits_{n =
0}^\infty {d_n r^{s_v + n}}
\end{equation}

\noindent
so that
\begin{equation}
\label{eq20}
\begin{array}{l}
 \frac{du}{dr} = \sum\limits_{n = 0}^\infty {\left( {s_u + n} \right)b_n
r^{s_u + n - 1}} \\
 \frac{dv}{dr} = \sum\limits_{n = 0}^\infty {\left( {s_v + n} \right)d_n
r^{s_v + n - 1}} \\
 \frac{d^2u}{dr^2} = \sum\limits_{n = 0}^\infty {\left( {s_u + n}
\right)\left( {s_u + n - 1} \right)b_n r^{s_u + n - 2}} \\
 \frac{d^2v}{dr^2} = \sum\limits_{n = 0}^\infty {\left( {s_v + n}
\right)\left( {s_v + n - 1} \right)d_n r^{s_v + n - 2}} \\
 \end{array}
\end{equation}

\noindent
substituting (\ref{eq19}) and (\ref{eq20}) into the equations (\ref{eq17}), we have
\begin{equation}
\label{eq21}
\begin{array}{l}
 \sum\limits_{n = 0}^\infty {\left\{ {\begin{array}{l}
 \left[ {\left( {s_u + n} \right)\left( {s_u + n - 1} \right) - \gamma
\left( {\gamma + 1} \right)} \right]b_n \\
 - 2\left[ {\frac{\sqrt {1 - \varepsilon _u^2 } }{\lambda }\left( {s_u + n -
1} \right) + \left( {Z\varepsilon _u + \mu } \right)} \right]b_{n - 1} \\
 \end{array}} \right\}r^{s_u + n - 2}} = 0 \\
 \sum\limits_{n = 0}^\infty {\left\{ {\begin{array}{l}
 \left[ {\left( {s_v + n} \right)\left( {s_v + n - 1} \right) - \gamma
\left( {\gamma - 1} \right)} \right]d_n \\
 - 2\left[ {\frac{\sqrt {1 - \varepsilon _v^2 } }{\lambda }\left( {s_v + n -
1} \right) + \left( {Z\varepsilon _v + \mu } \right)} \right]d_{n - 1} \\
 \end{array}} \right\}r^{s_v + n - 2}} = 0 \\
 \end{array}
\end{equation}

\noindent
finally we obtain the two recursive relation for the power series (\ref{eq19})
\begin{equation}
\label{eq22}
\begin{array}{l}
 \left[ {\left( {s_u + n} \right)\left( {s_u + n - 1} \right) - \gamma
\left( {\gamma + 1} \right)} \right]b_n - 2\left[ {\frac{\sqrt {1 -
\varepsilon _u^2 } }{\lambda }\left( {s_u + n - 1} \right) + \left(
{Z\varepsilon _u + \mu } \right)} \right]b_{n - 1} = 0 \\
 \left[ {\left( {s_v + n} \right)\left( {s_v + n - 1} \right) - \gamma
\left( {\gamma - 1} \right)} \right]d_n - 2\left[ {\frac{\sqrt {1 -
\varepsilon _v^2 } }{\lambda }\left( {s_v + n - 1} \right) + \left(
{Z\varepsilon _v + \mu } \right)} \right]d_{n - 1} = 0 \\
 \end{array}
\end{equation}

\noindent
using the initial value condition $b_{ - 1} = b_{ - 2} = \cdots = 0$, $d_{ -
1} = d_{ - 2} = \cdots = 0$and $b_0 \ne 0$, $d_0 \ne 0$, let $n = 0$ and
substitute it into the above recursive relation, it educes that
\begin{equation}
\label{eq23}
\begin{array}{l}
 s_u \left( {s_u - 1} \right) - \gamma \left( {\gamma + 1} \right) = 0 \\
 s_v \left( {s_v - 1} \right) - \gamma \left( {\gamma - 1} \right) = 0 \\
 \end{array}
\end{equation}

\noindent
this gives
\begin{equation}
\label{eq24}
s_{u1} = 1 + \gamma ,\quad s_{u2} = - \gamma ;\quad s_{v1} = \gamma ,\quad
s_{v2} = 1 - \gamma
\end{equation}

\noindent
in order to that the whole solutions (\ref{eq2}) of the original equation satisfy
the boundary condition, we have to choose
\begin{equation}
\label{eq25}
s_{u1} = 1 + \gamma ,\quad s_{v1} = \gamma
\end{equation}

\noindent
hence
\begin{equation}
\label{eq26}
u = \sum\limits_{n = 0}^\infty {b_n r^{1 + \gamma + n}} ,\quad v =
\sum\limits_{n = 0}^\infty {d_n r^{\gamma + n}}
\end{equation}

\noindent
and
\begin{equation}
\label{eq27}
\phi \left( r \right) = e^{ - \frac{\sqrt {1 - \varepsilon _u^2 } }{\lambda
}r}\sum\limits_{n = 0}^\infty {b_n r^{1 + \gamma + n}} ,\quad \psi \left( r
\right) = e^{ - \frac{\sqrt {1 - \varepsilon _v^2 } }{\lambda
}r}\sum\limits_{n = 0}^\infty {d_n r^{\gamma + n}}
\end{equation}

\noindent
the power series must be cut off so that the whole wave function is limit at
$r \to \infty $. It is assumed that $b_{n_r } \ne 0,\quad d_{n_r } \ne 0$
and $b_{n_r + 1} = b_{n_r + 2} = \cdots = 0$, $d_{n_r + 1} = d_{n_r + 2} =
\cdots = 0$. According to the recursive relation (\ref{eq22}), let $n = n_r + 1$, we
obtain
\begin{equation}
\label{eq28}
\begin{array}{l}
 - \left[ {\frac{\sqrt {1 - \varepsilon _u^2 } }{\lambda }\left( {s_u + n_r
} \right) + \left( {Z\varepsilon _u + \mu } \right)} \right]b_{n_r } = 0 \\
 - \left[ {\frac{\sqrt {1 - \varepsilon _v^2 } }{\lambda }\left( {s_v + n_r
} \right) + \left( {Z\varepsilon _v + \mu } \right)} \right]d_{n_r } = 0 \\
 \end{array}
\end{equation}

\noindent
it requires that
\begin{equation}
\label{eq29}
\begin{array}{l}
 \frac{\sqrt {1 - \varepsilon _u^2 } }{\lambda }\left( {1 + \gamma + n_r }
\right) + \left( {Z\varepsilon _u + \mu } \right) = 0 \\
 \frac{\sqrt {1 - \varepsilon _v^2 } }{\lambda }\left( {\gamma + n_r }
\right) + \left( {Z\varepsilon _v + \mu } \right) = 0 \\
 \end{array}
\end{equation}

\noindent
if we think little of, we would obtain the formal solution
\begin{equation}
\label{eq30}
\begin{array}{l}
 \varepsilon _{u1} = \frac{ - \lambda ^2Z\mu + \sqrt {\left( {1 + \gamma +
n_r } \right)^4 + \lambda ^2Z^2\left[ {\lambda ^2\mu ^2 + \left( {1 + \gamma
+ n_r } \right)^2} \right]} }{\lambda ^2Z^2 + \left( {1 + \gamma + n_r }
\right)^2} \\
 \varepsilon _{u2} = \frac{ - \lambda ^2Z\mu - \sqrt {\left( {1 + \gamma +
n_r } \right)^4 + \lambda ^2Z^2\left[ {\lambda ^2\mu ^2 + \left( {1 + \gamma
+ n_r } \right)^2} \right]} }{\lambda ^2Z^2 + \left( {1 + \gamma + n_r }
\right)^2} \\
 \varepsilon _{v1} = \frac{ - \lambda ^2Z\mu + \sqrt {\left( {\gamma + n_r }
\right)^4 + \lambda ^2Z^2\left[ {\lambda ^2\mu ^2 + \left( {\gamma + n_r }
\right)^2} \right]} }{\lambda ^2Z^2 + \left( {\gamma + n_r } \right)^2} \\
 \varepsilon _{v2} = \frac{ - \lambda ^2Z\mu - \sqrt {\left( {\gamma + n_r }
\right)^4 + \lambda ^2Z^2\left[ {\lambda ^2\mu ^2 + \left( {\gamma + n_r }
\right)^2} \right]} }{\lambda ^2Z^2 + \left( {\gamma + n_r } \right)^2} \\
 \end{array}
\end{equation}

in form, these results as the inevitable deductions of the second-order
differential equations (\ref{eq14}) include omnifarious logic problems.

\noindent
a) The solutions (\ref{eq30}) are different from the formula (\ref{eq12}). It shows that the
original formula (\ref{eq12}) is incorrect for the second-order differential
equations (\ref{eq4}).

b) For the same quantum system described by the equations (\ref{eq14}), the four
eigenvalues sets of the energy levels disobey the uniqueness of the solution
of the differential equations. Which eigenvalues set is correct?

c) Is the relativistic energy the positive number or the negative number? If
we delete the negative energy solution, we also have two eigenvales set
corresponding the positive energy. They also disobey the uniqueness of the
eignesolutions set for the differential equation.

d) It is the most serious that, the solutions (\ref{eq30}) implying the (\ref{eq12}) are the
formal solution. Because of the definition $\gamma = \frac{\left| \kappa
\right|}{\kappa }\sqrt {\kappa ^2 + \lambda ^2\left( {\mu ^2 - Z^2} \right)}
$ given in the original article, only when $\kappa < 0$, the quadratic
equations with one unknown can have the solutions (\ref{eq30}). Whereas the $\kappa
$ values constructed by Dirac are actually $\pm 1,\;\pm 2,\; \cdots $, also
given in the original articles.

All of these problems are the mathematical and physical contradictions.
Consequently, we don't think the original solution (\ref{eq12}) is not deceitful
solution.

\section{Schr\"{o}dinger-like equation (\ref{eq10}) is bogus}

We even doubt that the second-order equation (\ref{eq10}) is the correct deduction
of the original coupled first-order equation (\ref{eq4}). By all appearances, from
(\ref{eq4}) to (\ref{eq10}), it actually introduces the transformation (\ref{eq8}) with (\ref{eq9}), namely
\begin{equation}
\label{eq31}
\left( {{\begin{array}{*{20}c}
 {\frac{\lambda }{C + \varepsilon }\left( {\frac{S}{\lambda } - \frac{\gamma
}{r} + \frac{d}{dr}} \right)\phi ^ - } \hfill \\
 {\frac{\lambda }{C - \varepsilon }\left( { - \frac{S}{\lambda } +
\frac{\gamma }{r} + \frac{d}{dr}} \right)\phi ^ + } \hfill \\
\end{array} }} \right) = \left( {{\begin{array}{*{20}c}
 {\cos \frac{\lambda \eta }{2}} \hfill & {\sin \frac{\lambda \eta }{2}}
\hfill \\
 { - \sin \frac{\lambda \eta }{2}} \hfill & {\cos \frac{\lambda \eta }{2}}
\hfill \\
\end{array} }} \right)\left( {{\begin{array}{*{20}c}
 g \hfill \\
 f \hfill \\
\end{array} }} \right)
\end{equation}

\noindent
it gives the separate form
\begin{equation}
\label{eq32}
\begin{array}{l}
 g = \frac{\frac{\lambda }{C + \varepsilon }\left( {\frac{S}{\lambda } -
\frac{\gamma }{r} + \frac{d}{dr}} \right)\phi ^ - - \frac{\lambda }{C -
\varepsilon }\left( { - \frac{S}{\lambda } + \frac{\gamma }{r} +
\frac{d}{dr}} \right)\phi ^ + }{\cos \frac{\lambda \eta }{2} + \sin
\frac{\lambda \eta }{2}} \\
 f = \frac{\frac{\lambda }{C + \varepsilon }\left( {\frac{S}{\lambda } -
\frac{\gamma }{r} + \frac{d}{dr}} \right)\phi ^ - + \frac{\lambda }{C -
\varepsilon }\left( { - \frac{S}{\lambda } + \frac{\gamma }{r} +
\frac{d}{dr}} \right)\phi ^ + }{\cos \frac{\lambda \eta }{2} + \sin
\frac{\lambda \eta }{2}} \\
 \end{array}
\end{equation}

\noindent
these transformations cannot translate the equations (\ref{eq4}) namely
\begin{equation}
\label{eq33}
\begin{array}{l}
 \left( {1 + \lambda ^2\frac{Z + \mu }{r} - \varepsilon } \right)g\left( r
\right) + \lambda \left( {\frac{k}{r} - \frac{d}{dr}} \right)f\left( r
\right) = 0 \\
 \lambda \left( {\frac{k}{r} + \frac{d}{dr}} \right)g\left( r \right) -
\left( { - 1 + \lambda ^2\frac{Z - \mu }{r} - \varepsilon } \right)f\left( r
\right) = 0 \\
 \end{array}
\end{equation}

\noindent
into any Schr\"{o}dinger-like second-order equations, and the second-order
equation (\ref{eq10}) give in the original paper is not correct.

\section{Correct solution of the corresponding first-order equation}

Because there has been not any experiment data to approve the imagination,
we don't think that the mass of electron in the Coulomb field is really
dependent on position. Consequently, in principle, it should be meaningles
to find the correct solution of the equation (\ref{eq4}) or (\ref{eq33}). When only looking
from a mathematical point of view, we give the correct exact solution of the
equations (\ref{eq4}) or (\ref{eq33}). It is well known that directly solving the original
coupled first-order equation (\ref{eq4}) namely (\ref{eq33}) is simpler out and away than
translating it into the so-called Schr\"{o}dinger-like equation to obtain
the exact solution. One firstly note the behavior of $f\left( r \right)$ and
$g\left( r \right)$ for $g \to \infty $, since neglecting the terms
proportional to $1 \mathord{\left/ {\vphantom {1 r}} \right.
\kern-\nulldelimiterspace} r$ the differential equations (\ref{eq33}) read
\begin{equation}
\label{eq34}
\left( {1 - \varepsilon } \right)g - \lambda \frac{df}{dr} = 0,\quad \lambda
\frac{dg}{dr} + \left( {1 + \varepsilon } \right)f = 0
\end{equation}

\noindent
it follows immediately that
\begin{equation}
\label{eq35}
\frac{d^2f}{dr^2} + \frac{1 - \varepsilon ^2}{\lambda ^2}f \sim 0,\quad
\frac{d^2g}{dr^2} + \frac{1 - \varepsilon ^2}{\lambda ^2}g \sim 0
\end{equation}

\noindent
there will be different solution with the different fields of definitions of
the $\varepsilon $. However, it is considered all along that the
relativistic energy in the Coulomb field satisfy the condition $0 < E < m_0
c^2$ namely $E < 1$. According to (\ref{eq13}) and (\ref{eq14}), it only gives $1 <
\varepsilon ^2 < 2\lambda ^2 + 1$. For the moment, it is considered that
$\varepsilon ^2 > 1$. We obtain the asymptotic solutions of the equations
(\ref{eq33})
\begin{equation}
\label{eq36}
f \sim e^{ - \frac{\sqrt {\varepsilon ^2 - 1} }{\lambda }r},\quad g \sim e^{
- \frac{\sqrt {\varepsilon ^2 - 1} }{\lambda }r}
\end{equation}

\noindent
and the exact solution of the equations (\ref{eq33}) take form
\begin{equation}
\label{eq37}
f = e^{ - \frac{\sqrt {\varepsilon ^2 - 1} }{\lambda }r}u,\quad g = e^{ -
\frac{\sqrt {\varepsilon ^2 - 1} }{\lambda }r}v
\end{equation}

\noindent
they give
\begin{equation}
\label{eq38}
\begin{array}{l}
 \frac{df}{dr} = e^{ - \frac{\sqrt {\varepsilon ^2 - 1} }{\lambda
}r}\frac{du}{dr} - \frac{\sqrt {\varepsilon ^2 - 1} }{\lambda }e^{ -
\frac{\sqrt {\varepsilon ^2 - 1} }{\lambda }r}u \\
 \frac{dg}{dr} = e^{ - \frac{\sqrt {\varepsilon ^2 - 1} }{\lambda
}r}\frac{dv}{dr} - \frac{\sqrt {\varepsilon ^2 - 1} }{\lambda }e^{ -
\frac{\sqrt {\varepsilon ^2 - 1} }{\lambda }r}v \\
 \end{array}
\end{equation}

\noindent
substituting for equations (\ref{eq33}), we have
\begin{equation}
\label{eq39}
\begin{array}{l}
 \lambda \frac{du}{dr} - \left( {\frac{\lambda k}{r} + \sqrt {\varepsilon ^2
- 1} } \right)u + \left[ {\left( {\varepsilon - 1} \right) - \frac{\lambda
^2\left( {Z + \mu } \right)}{r}} \right]v = 0 \\
 \lambda \frac{dv}{dr} + \left( {\frac{\lambda k}{r} - \sqrt {\varepsilon ^2
- 1} } \right)v + \left[ {\left( {\varepsilon + 1} \right) - \frac{\lambda
^2\left( {Z - \mu } \right)}{r}} \right]u = 0 \\
 \end{array}
\end{equation}

\noindent
finding the power series solution, put
\begin{equation}
\label{eq40}
v = \sum\limits_{n = 0}^\infty {b_n } r^{\sigma + n},\quad u =
\sum\limits_{n = 0}^\infty {d_n } r^{\sigma + n}
\end{equation}

\noindent
substitute it into the above equations, we obtain
\begin{equation}
\label{eq41}
\begin{array}{l}
 \lambda \sum\limits_{n = 0}^\infty {\left( {\sigma + n} \right)d_n }
r^{\sigma + n - 1} - \left( {\frac{\lambda k}{r} + \sqrt {\varepsilon ^2 -
1} } \right)\sum\limits_{n = 0}^\infty {d_n } r^{\sigma + n} + \left[
{\left( {\varepsilon - 1} \right) - \frac{\lambda ^2\left( {Z + \mu }
\right)}{r}} \right]\sum\limits_{n = 0}^\infty {b_n } r^{\sigma + n} = 0 \\
 \lambda \sum\limits_{n = 0}^\infty {\left( {\sigma + n} \right)b_n }
r^{\sigma + n - 1} + \left( {\frac{\lambda k}{r} - \sqrt {\varepsilon ^2 -
1} } \right)\sum\limits_{n = 0}^\infty {b_n } r^{\sigma + n} + \left[
{\left( {\varepsilon + 1} \right) - \frac{\lambda ^2\left( {Z - \mu }
\right)}{r}} \right]\sum\limits_{n = 0}^\infty {d_n } r^{\sigma + n} = 0 \\
 \end{array}
\end{equation}

\noindent
it predigests that
\begin{equation}
\label{eq42}
\begin{array}{l}
 \sum\limits_{n = 0}^\infty {\left[ {\lambda \left( {\sigma + n - k}
\right)d_n - \sqrt {\varepsilon ^2 - 1} d_{n - 1} + \left( {\varepsilon - 1}
\right)b_{n - 1} - \lambda ^2\left( {Z + \mu } \right)b_n } \right]}
r^{\sigma + n - 1} = 0 \\
 \sum\limits_{n = 0}^\infty {\left[ {\lambda ^2\left( {Z - \mu } \right)d_n
- \left( {\varepsilon + 1} \right)d_{n - 1} - \lambda \left( {\sigma + n +
k} \right)b_n + \sqrt {\varepsilon ^2 - 1} b_{n - 1} } \right]} r^{\sigma +
n - 1} = 0 \\
 \end{array}
\end{equation}

\noindent
so the coefficients of the power series satisfy the recursive relations
\begin{equation}
\label{eq43}
\begin{array}{l}
 \lambda \left( {\sigma + n - k} \right)d_n - \sqrt {\varepsilon ^2 - 1}
d_{n - 1} + \left( {\varepsilon - 1} \right)b_{n - 1} - \lambda ^2\left( {Z
+ \mu } \right)b_n = 0 \\
 \lambda ^2\left( {Z - \mu } \right)d_n - \left( {\varepsilon + 1}
\right)d_{n - 1} - \lambda \left( {\sigma + n + k} \right)b_n + \sqrt
{\varepsilon ^2 - 1} b_{n - 1} = 0 \\
 \end{array}
\end{equation}

Solving the above recursive relations and using the initial conditions that
$b_0 \ne 0$, $b_{ - 1} = b_{ - 2} = \cdots = 0$ and $d_0 = 0$, $d_{ - 1} =
d_{ - 2} = \cdots = 0$, put $n = 0$ in (\ref{eq43}), we obtain
\begin{equation}
\label{eq44}
\begin{array}{l}
 \lambda \left( {\sigma - k} \right)d_0 - \lambda ^2\left( {Z + \mu }
\right)b_0 = 0 \\
 \lambda ^2\left( {Z - \mu } \right)d_0 - \lambda \left( {\sigma + k}
\right)b_0 = 0 \\
 \end{array}
\end{equation}

\noindent
it requests that the determinant of coefficient is equivalent to zero, so
that
\begin{equation}
\label{eq45}
\left| {{\begin{array}{*{20}c}
 {\lambda \left( {\sigma - k} \right)} \hfill & { - \lambda ^2\left( {Z +
\mu } \right)} \hfill \\
 {\lambda ^2\left( {Z - \mu } \right)} \hfill & { - \lambda \left( {\sigma +
k} \right)} \hfill \\
\end{array} }} \right| = 0
\end{equation}

\noindent
it educes that value of the index $\sigma = \pm \sqrt {k^2 + \lambda
^2\left( {Z^2 - \mu ^2} \right)} $, since the wave function must be limited
at $r \to 0$, it can be only taken
\begin{equation}
\label{eq46}
\sigma = \sqrt {k^2 + \lambda ^2\left( {Z^2 - \mu ^2} \right)}
\end{equation}

\noindent
on the other hand, the formal whole wave function should be combined by
expressions (\ref{eq2}), (\ref{eq3}), (\ref{eq37}), (\ref{eq40}) and (\ref{eq46}), that is to say
\begin{equation}
\label{eq47}
\psi = \left( {{\begin{array}{*{20}c}
 {ie^{ - \frac{\sqrt {\varepsilon ^2 - 1} }{\lambda }r}\sum\limits_{n =
0}^\infty {b_n } r^{\sqrt {k^2 + \lambda ^2\left( {Z^2 - \mu ^2} \right)} +
n - 1}\chi _{\iota m}^j } \hfill \\
 {e^{ - \frac{\sqrt {\varepsilon ^2 - 1} }{\lambda }r}\sum\limits_{n =
0}^\infty {d_n } r^{\sqrt {k^2 + \lambda ^2\left( {Z^2 - \mu ^2} \right)} +
n - 1}\vec {\sigma } \cdot \chi _{\iota m}^j } \hfill \\
\end{array} }} \right)
\end{equation}

\noindent
the boundary condition at $r \to \infty $ requests that the parts of the
power series in the formal whole wave function must be cut off. It is
assumed that $b_{n_r } \ne 0,\quad d_{n_r } \ne 0$ and $b_{n_r + 1} = b_{n_r
+ 2} = \cdots = 0$, $d_{n_r + 1} = d_{n_r + 2} = \cdots = 0$, according to
the recursive relations (\ref{eq43}), put $n = n_r + 1$, we obtain
\begin{equation}
\label{eq48}
\begin{array}{l}
 - \sqrt {\varepsilon ^2 - 1} d_{n_r } + \left( {\varepsilon - 1}
\right)b_{n_r } = 0 \\
 - \left( {\varepsilon + 1} \right)d_{n_r } + \sqrt {\varepsilon ^2 - 1}
b_{n_r } = 0 \\
 \end{array}
\end{equation}

\noindent
it indicates that the $\varepsilon $ should take some special values so that
$b_{n_r + 1} = 0$ and $d_{n_r + 1} = 0$. Note that the recursive relations
(\ref{eq43}). Multiplying the first relation by $\sqrt {\varepsilon ^2 - 1} $ and
multiplying the second relation by $\varepsilon - 1$, we have
\begin{equation}
\label{eq49}
\begin{array}{l}
 \lambda \left( {\sigma + n_r - k} \right)\sqrt {\varepsilon ^2 - 1} d_{n_r
} - \left( {\varepsilon ^2 - 1} \right)d_{n_r - 1} \\
 - \lambda ^2\left( {Z + \mu } \right)\sqrt {\varepsilon ^2 - 1} b_{n_r } +
\left( {\varepsilon - 1} \right)\sqrt {\varepsilon ^2 - 1} b_{n_r - 1} = 0
\\
 \lambda ^2\left( {Z - \mu } \right)\left( {\varepsilon - 1} \right)d_{n_r }
- \left( {\varepsilon ^2 - 1} \right)d_{n_r - 1} \\
 - \lambda \left( {\sigma + n_r + k} \right)\left( {\varepsilon - 1}
\right)b_{n_r } + \left( {\varepsilon - 1} \right)\sqrt {\varepsilon ^2 - 1}
b_{n_r - 1} = 0 \\
 \end{array}
\end{equation}

\noindent
this deduces that
\begin{equation}
\label{eq50}
d_{n_r } = - \frac{\left( {\sigma + n_r + k} \right)\left( {\varepsilon - 1}
\right) - \lambda \left( {Z + \mu } \right)\sqrt {\varepsilon ^2 - 1}
}{\left( {\sigma + n_r - k} \right)\sqrt {\varepsilon ^2 - 1} - \lambda
\left( {Z - \mu } \right)\left( {\varepsilon - 1} \right)}b_{n_r }
\end{equation}

\noindent
combing this relations and one of the system of the recursive relations
(\ref{eq48}), we obtain
\begin{equation}
\label{eq51}
\left( {\sigma + n_r } \right)\sqrt {\varepsilon ^2 - 1} = \lambda
Z\varepsilon + \lambda \mu
\end{equation}

\noindent
it gives that
\begin{equation}
\label{eq52}
\varepsilon = \frac{\lambda ^2\mu Z\pm \left( {\sigma + n_r } \right)\sqrt
{\left( {\sigma + n_r } \right)^2 + \lambda ^2\left( {\mu ^2 - Z^2} \right)}
}{\left( {\sigma + n_r } \right)^2 - \lambda ^2Z^2}
\end{equation}

\noindent
we find that it is different from not only the formula (\ref{eq11}) given in the
original articles but also the correct formula (\ref{eq30}) of the second-order
differential equations. This shows that the original exact solution of the
Dirac equation with position-dependent mass is the pseudo solution.

\section{Conclusions}

We have used the basal knowledge of the differential equation with variable
coefficient to show that, almost every step of the mathematical operation in
the mentioned paper for introducing the Dirac theory with the
position-dependent mass of the electron in the Coulomb field is incorrect,
and the energy eigenvalues so the eigen-wave-function given in the original
paper are pseudo. Only in a mathematical signification, we give the correct
solution and eigenvalue set of the first-order differential equations.
However we don't regard the formula (\ref{eq52}) of the energy-levels as the
necessary result of the development of quantum mechanics. Because the
position-dependent mass of the electron in the Coulomb field make us
discredit its authenticity.

Some other papers also alleged that they found the exact solution of the
Dirac equation for a particle with position-dependent
mass\cite{Vakarchuk:2005}, which might be useful in the study of the
corresponding non-relativistic problem as a reference result. It was even
considered that the next terms, which they have neglected in this work (in
particular the dipolar one) and which are responsible for the super-fine
structure of the energy spectrum, can be taken into account by means of
standard perturbation theory. Although the corresponding paper constructed
the second-order Dirac equation by using unconventionality methods which
also can be seen in the some papers published 25 years ago\cite{Wong:1982}.
However, is this theory correct? Without solving the corresponding
differential equation, one can find some methods to directly conclude that
some other papers about the Dirac theory are not correct.

\end{document}